\documentclass[useAMS,usegraphicx,usenatbib]{mn2e}

\newcommand{\Msun}{\ensuremath{\,{\rm M}_\odot}}           
\newcommand{\Rsun}{\ensuremath{\,{\rm R}_\odot}}           
\newcommand{\Teff}{\ensuremath{T_{\rm eff}}}               
\newcommand{\Mjup}{\ensuremath{\,{\rm M}_{\rm Jup}}}       
\newcommand{\Rjup}{\ensuremath{\,{\rm R}_{\rm Jup}}}       
\newcommand{\kms}{\,km\,s$^{-1}$}                          
\newcommand{\mss}{\,m\,s$^{-2}$}                           
\newcommand{\mc}[1]{\multicolumn{2}{c}{#1}}
\newcommand{\as}{\ensuremath{^{\prime\prime}}}              
\newcommand{\am}{\ensuremath{^\prime}}                      
\newcommand{\er}[3]{\ensuremath{#1^{+#2}_{-#3}}}
\newcommand{\erm}[3]{\mc{\ensuremath{#1^{+#2}_{-#3}}}}
\newcommand{\MoH}{\ensuremath{\left[\frac{\rm M}{\rm H}\right]}}        
\newcommand{\pjup}{\ensuremath{\,\rho_{\rm Jup}}}          
\newcommand{\psun}{\ensuremath{\,\rho_\odot}}              
\newcommand{\reff}[1]{{#1}}                                  

\title[High-precision defocussed photometry of WASP-5]
      {High-precision photometry by telescope defocussing. I. The transiting planetary system WASP-5\thanks{Based on data collected by MiNDSTEp with the Danish 1.54\,m telescope at the ESO La Silla Observatory}}

\author[Southworth et al.]
       {John Southworth\,$^1$%
        \thanks{E-mail: j.k.taylor@warwick.ac.uk},
        T.\ C.\ Hinse\,$^{2,3}$,          
        U.\ G.\ J{\o}rgensen\,$^2$,       
        M.\ Dominik\,$^4$\thanks{Royal Society University Research Fellow},                
        D.\ Ricci\,$^5$,                  
        \newauthor
        M.\ J.\ Burgdorf\,$^6$,           
        A.\ Hornstrup\,$^7$,              
        P.\ J.\ Wheatley\,$^1$,           
        T.\ Anguita\,$^8$,                
        V.\ Bozza\,$^{9,10}$,             
        \newauthor
        S.\ Calchi Novati\,$^{9,10}$,     
        K.\ Harps{\o}e\,$^2$,             
        P.\ Kj{\ae}rgaard\,$^2$,          
        C.\ Liebig\,$^8$,                 
        L.\ Mancini\,$^{9,10}$,           
        \newauthor
        G.\ Masi\,$^{11}$,                
        M.\ Mathiasen\,$^2$,              
        S.\ Rahvar\,$^{12}$,              
        G.\ Scarpetta\,$^{9,10}$,         
        C.\ Snodgrass\,$^{13}$,           
        \newauthor
        J.\ Surdej\,$^5$,                 
        C.\ C.\ Th\"one$^{14,15}$,        
        M.\ Zub$^8$                       
        \\
        $^1$\,Department of Physics, University of Warwick, Coventry, CV4 7AL, UK \\
        $^2$\,Niels Bohr Institute, University of Copenhagen, Juliane Maries Vej 30, Copenhagen \O, 2100, Denmark \\
        $^3$\,Armagh Observatory, College Hill, Armagh, BT61 9DG, Northern Ireland, UK \\
        $^4$\,SUPA, University of St Andrews, School of Physics \& Astronomy, North Haugh, St Andrews, KY16 9SS, UK \\
        $^5$\,Institut d'Astrophysique et de G\'eophysique, Universit\'e de Li\`ege, 4000 Li\`ege, Belgium \\
        $^6$\,Astrophysics Research Institute, Liverpool John Moores University,
                                          Twelve Quays House, Egerton Wharf, Birkenhead, CH41 1LD, UK \\
        $^7$\,National Space Institute, Technical University of Denmark,
                                          Juliane Maries Vej 30, Copenhagen \O, 2100, Denmark  \\
        $^8$\,Astronomisches Rechen-Institut, Zentrum f\"ur Astronomie, Universit\"at Heidelberg,
                                          M\"onchhofstrasse 12-14, 69120 Heidelberg, Germany \\
        $^9$\,Dipartimento di Fisica ``E. R. Caianiello'', Universit\`a di Salerno, Baronissi, Italy \\
        $^{10}$\,Istituto Nazionale di Fisica Nucleare, Sezione di Napoli, Italy \\
        $^{11}$\,Bellatrix Observatory, Centre for Backyard Astrophysics, Ceccano (FR), Italy \\
        $^{12}$\,Sharif University of Technology, Tehran, Iran \\
        $^{13}$\,European Southern Observatory, Casilla 19001, Santiago 19, Chile \\
        $^{14}$\,Dark Cosmology Centre, Niels Bohr Institute, University of Copenhagen,
                                          Juliane Maries Vej 30, Copenhagen \O, 2100, Denmark \\
        $^{15}$\,INAF, Osservatorio Astronomico di Brera, 23807 Merate, Italy
        }

\begin{document} \maketitle 

\begin{abstract}
We present high-precision photometry of two transit events of the extrasolar planetary system WASP-5, obtained with the Danish 1.54\,m telescope at ESO La Silla. \reff{In order to minimise both random and flat-fielding errors, we defocussed the telescope so its point spread function approximated an annulus of diameter 40 pixels (16\as). Data reduction was undertaken using standard aperture photometry plus an algorithm for optimally combining the ensemble of comparison stars. The resulting light curves have point-to-point scatters of 0.50\,mmag for the first transit and 0.59\,mmag for the second. We construct detailed signal to noise calculations for defocussed photometry, and apply them to our observations. We model the light curves with the {\sc jktebop} code and combine the results with tabulated predictions from theoretical stellar evolutionary models to derive the physical properties of the WASP-5 system. We find that the planet has a mass of $M_{\rm b} = 1.637 \pm 0.075 \pm 0.033$\Mjup, a radius of $R_{\rm b} = 1.171 \pm 0.056 \pm 0.012$\Rjup, a large surface gravity of $g_{\rm b} = 29.6 \pm 2.8$\mss\ and a density of $\rho_{\rm b} = 1.02 \pm 0.14 \pm 0.01$\pjup\ (statistical and systematic uncertainties). The planet's high equilibrium temperature of $T_{\rm eq} = 1732 \pm 80$\,K, makes it a good candidate for detecting secondary eclipses.}
\end{abstract}



\section{Introduction}

The discovery of the first extrasolar planet \citep{MayorQueloz95nat} opened a new field of astronomical research. We now know of over 300 extrasolar planets%
\footnote{See {\tt http://exoplanet.eu/} for a list of known extrasolar planets.}%
, which encourages statistical studies of these objects \citep[e.g.][]{UdrySantos07araa}. However, most of these discoveries have been made via measurements of the radial velocities of their parent stars, which yield relatively little information about individual planets \reff{(only the orbital parameters and lower limits on the planet's mass)}.

The discovery of transiting extrasolar planets (hereafter TEPs) in the year 1999 \citep{Charbonneau+00apj,Henry+00apj} has provided a new window on the properties of extrasolar gas-giant planets, making it possible to derive the physical properties of planets and their parent stars (with a little help from stellar theory). These analyses require high-precision photometry of planetary transits in order to measure the radii of the components and the inclination of their orbital plane with respect to Earth.

Good light curves have been obtained for a substantial number of the roughly fifty known TEPs, but the process remains fraught with difficulties. Ground-based studies are subject to complications arising due to scintillation, atmospheric effects which change throughout transit observations, telescope tracking errors, and problems with saturation and flat-fielding of the ubiquitous charge-coupled device (CCD) detectors. \reff{Space-based facilities, however, can be subject to major cost and calibration issues.}

\reff{We have therefore decided to experiment with heavy defocussing of ground-based telescopes in order to nullify as many as possible of the above effects. The idea is to choose relatively long integration times (several minutes) and disperse the resulting large numbers of photons over many CCD pixels. This has the huge advantage that flat-fielding errors can be averaged down by orders of magnitude compared to focussed observations, and also means that normal changes in atmospheric seeing are irrelevant. Nonlinear pixel responses should be corrected by calibrating the response of the CCD or avoided by sticking to count levels where such effects are low. The longer integration times and large point spread functions (PSFs) mean that the sky background level is much higher than for standard approaches, but signal-to-noise calculations (Section\,\ref{sec:sn}) show that this is unimportant in many cases.}

In this work we present high-precision photometry of WASP-5, a $V=12.3$ star which was discovered to harbour a TEP by \citet{Anderson+08mn}. We used the Danish 1.54\,m telescope at ESO La Silla, Chile, as part of the 2008 MiNDSTEp campaign%
\footnote{\reff{Information on the MiNDSTEp (Microlensing Network for the Detection of Small Terrestrial Exoplanets) campaign can be found at {\tt www.mindstep-science.org/}}}%
, and the DFOSC imaging camera. \reff{It was suspected that DFOSC would not be suited to these observations, as focal-reducing instruments can suffer internal reflections, but this does not appear to be a problem.} We observed two transit events and achieved observational scatters of only 0.50\,mmag and 0.59\,mmag per point in the final light curves. In the following sections we discuss our observations and data reduction procedures, and then analyse the data using the methods detailed by \citet{Me08mn,Me09mn}. Our results are fully homogeneous with those of the fourteen TEPs analysed in these papers, and represent the highest-precision measurements of the physical properties of the star and planet in the WASP-5 system.

\reff{Whilst it is unusual to break the 1\,mmag barrier in astronomical differential photometry, our observations do not set any records. \citet{Gilliland+93aj} obtained observational scatters of 0.25\,mmag per minute in time-series observations of stars in the open cluster M\,67. These authors used a set of 4m-class telescopes, allowing them to achieve a lower Poisson and scintillation noise than ourselves, and CCD imagers with much finer pixel scale, which meant much less defocussing and so a lower sky background.}

A number of researchers have used some telescope defocussing to improve the photometric accuracy of light curves of TEPs, \reff{in many cases} broadening the PSFs by only a few pixels. A partial list includes studies of GJ\,436 \citep{Gillon+07aa2,Demory+07aa,Alonso+08aa}, HD\,17156 \citep{Gillon+08aa}, HD\,189733 \citep{Winn+07aj}, HAT-P-1 \citep{Winn+07aj2}, {\it Hubble Space Telescope} observations of HD\,189733 \citep{Swain+08nat}, and the NASA EPOXI mission \citep{Christiansen+08xxx}. \reff{The discovery paper of WASP-5 \citep{Anderson+08mn} presented a light curve from a 1.2\,m telescope which achieved a scatter of about 1\,mmag using this technique. Substantial defocussing was also used by \citet{Gillon+08xxx} in the course of obtaining a light curve of WASP-5 using the VLT, but the resulting data suffer systematic effects which were not removable.} These were attributed to deformation of the primary mirror, as the active optics had to be turned off to permit defocussing. Finally, PSF broadening has been performed both by moving the telescope during individual exposures \citep{Bakos+04pasp} and by shuffling charge at will round an orthogonal-transfer CCD device \citep{Tonry+05pasp,Johnson+08xxx}.


\section{Observations}

We observed two transits of WASP-5 using the 1.54\,m Danish%
\footnote{Information on the 1.54\,m Danish Telescope and DFOSC can be found at
{\tt http://www.eso.org/sci/facilities/lasilla/telescopes/d1p5/}}
Telescope at ESO La Silla and the DFOSC focal-reducing imager. This setup yielded a full field of view of 13.7\am$\times$13.7\am\ and a plate scale of 0.39\as\,pixel$^{-1}$. For the second set of observations the CCD was windowed down to approximately 2000$\times$800 pixels to decrease the readout time \reff{from 91\,s to 39\,s}. All observations were done through the Cousins $R$ filter, which allowed a higher count rate than Johnson $V$ for red stars but did not suffer the fringing effects visible with DFOSC and the Cousins $I$ filter. An observing log is given in Table\,\ref{tab:obslog}. \reff{The observing conditions and sky transparency were good throughout both nights.}

\begin{figure} \includegraphics[width=0.48\textwidth,angle=0]{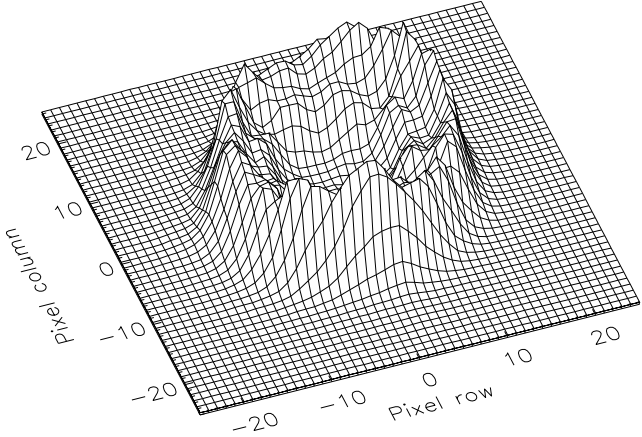}
\caption{\label{fig:psf} Surface plot of the PSF of WASP-5 in an image taken at
random from the observing sequence on the night of 2008 August 29th. The $x$ and
$y$ axes are in pixels. The lowest and highest counts are 623 and 28429 ADUs,
respectively, and the $z$ axis is on a linear scale.} \end{figure}

The exposure time was set at 120\,s, to allow a large amount of light to be collected in each observation whilst avoiding undersampling the light variations through transit \reff{(see Section\,\ref{sec:sn} for detailed signal to noise calculations)}. We then adjusted the amount of defocussing until the peak counts per pixel from WASP-5 were roughly 25\,000 above the sky background. This resulted in a doughnut-shaped PSF with a diameter of about 40 pixels (16\as) and containing roughly $10^7$ counts from WASP-5. Once the amount of defocussing was settled on, it was not changed until the end of the observing sequence. An example PSF of WASP-5 is shown in Fig.\,\ref{fig:psf}.

A few images were also taken with the telescope properly focussed, and were used to verify that there were no faint stars within the PSF of WASP-5 which might dilute the transit depth. We found that the closest detectable star to WASP-5 is at a distance of 44\as\ (113 pixels), so the edge of its PSF is separated from that of WASP-5 by over 70 pixels in our defocussed images. The closest stars of similar to or greater brightness than WASP-5 lie over 4\am\ away. We conclude that no stars interfere with the PSF of WASP-5.

\begin{table} \centering
\setlength{\tabcolsep}{3pt}
\caption{\label{tab:obslog} Log of the observations presented in this work.}
\begin{tabular}{lccccccc} \hline \hline
Date & Start time & End time & Number of    & Exposure & Airmass \\
     & (UT)       & (UT)     & observations & time (s) &        \\
\hline
2008 08 29 & 05:37 & 10:08 &  74 & 120.0 & 1.03--1.54 \\
2008 09 21 & 01:06 & 05:32 & 101 & 120.0 & 1.03--1.40 \\
\hline \hline \end{tabular} \end{table}

\section{Data reduction}

Our data reduction pipeline is written in the {\sc idl}%
\footnote{The acronym {\sc idl} stands for Interactive Data Language and is a trademark of ITT Visual
Information Solutions. For further details see {\tt http://www.ittvis.com/ProductServices/IDL.aspx}.}
programming language and uses the {\sc daophot} photometry package \citep{Stetson87pasp} (distributed as part of the {\sc astrolib}%
\footnote{The {\sc astrolib} subroutine library is distributed by NASA. For further details see {\tt http://idlastro.gsfc.nasa.gov/}.}
library) to perform aperture photometry. The telescope was autoguided thoughout our observations so it was possible to fix the positions of the circular apertures for each dataset. We experimented with different aperture radii, finding that this had a small effect on the scatter in the photometry but no detectable effect on the shape of the resulting transits. We chose the apertures which yielded the most precise photometry, with radii of 28 pixels for the PSF and 38 and 60 pixels for the inner and outer edges of the sky regions, respectively. Thus a PSF aperture covered 2463 pixels and a sky region covered 6773 pixels.

We used these aperture sizes and the {\sc astrolib}/{\sc aper} routine to obtain aperture photometry of all stars in our images with sufficient counts to be useful. The number of comparison stars was either eight or nine for these observations. The differential-magnitude light curves of all these stars were checked for short-timescale variability, and none was found. Most of the comparison star light curves have slow variations in their apparent brightness, attributable to atmospheric effects, so variability on timescales greater than about 5\,hr would not be detectable.

To remove these slow variations in the differential-magnitude light curves we implemented an algorithm to optimise the selection of comparison stars. We defined time intervals outside transit and used the {\sc idl}/{\sc amoeba} routine to minimise the square of the magnitude values within these intervals, resulting in light curves which were normalised to zero differential magnitude. The parameters of the minimisation were $N$ coefficients of a polynomial versus time, and weights of each comparison star relative to the first comparison star. All good comparison stars were combined into one ensemble by weighted flux summation. For each dataset we used $N = 2$, which removes overall variations without deforming the shape of the transit. We found that it was necessary to have $N > 1$ because we have too few comparison stars to create an ensemble with a low Poisson noise and the same photometric colour as the target star. We found that the best-fitting comparison star weights were usually close to unity, as expected for observations dominated by Poisson noise.

\begin{figure} \includegraphics[width=0.48\textwidth,angle=0]{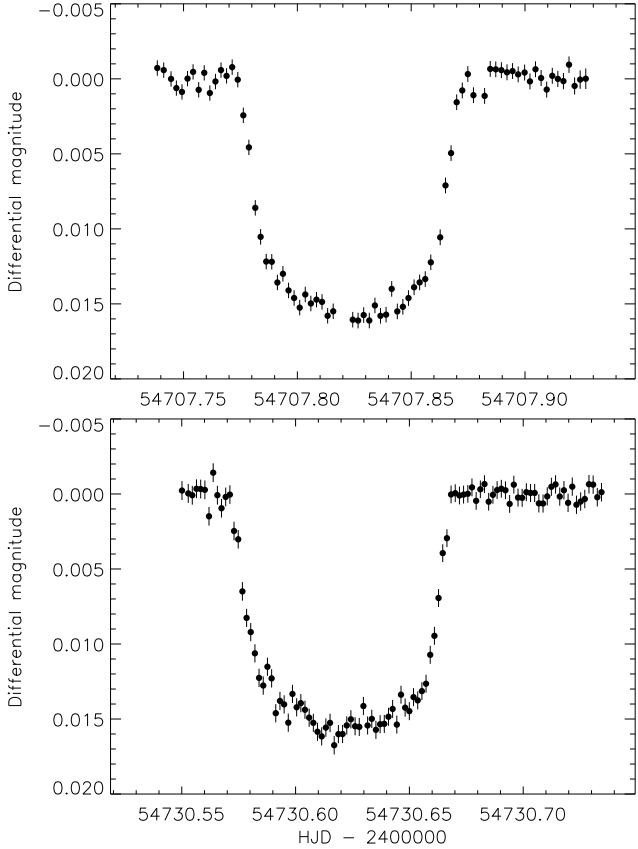}
\caption{\label{fig:plotlc} Final light curves of WASP-5 from our two
nights of observations. The observational error bars were too conservative
so have been shrunk to match the scatter of the data.} \end{figure}

Finally, we experimented with the inclusion of bias and flat-field corrections to the science images. Master bias and flat field images were created by median-combining sets of bias images and dome flats. We found that subtraction of a master bias image had a negligible effect on the photometry, as the bias level of the DFOSC CCD is well-behaved and exhibits minimal changes which are simply subsumed into the estimation of the sky background. Including a flat-fielding correction, however, gives a small but noticeable decrease in the scatter of the resulting light curves. The flat-fielding is relatively unimportant in this case because the stellar PSFs were kept on the same pixels by autoguiding the telescope. We have applied bias and flat-field corrections to generate our final light curves, which are shown in Fig.\,\ref{fig:plotlc} and reproduced in Table\,\ref{tab:lc}.

\begin{table} \centering
\caption{\label{tab:lc} Excerpts of the light curve of WASP-5. The full dataset
will be made available in the electronic version of this paper and at the CDS.}
\begin{tabular}{lrr} \hline \hline
HJD & Differential magnitude & Uncertainty \\
\hline
2454707.739980 & $-$0.000722 & 0.000512 \\
2454707.742816 & $-$0.000580 & 0.000513 \\
2454707.746034 &    0.000004 & 0.000512 \\
2454707.748360 &    0.000616 & 0.000512 \\
2454707.750814 &    0.000870 & 0.000513 \\[2pt]
2454730.554371 & $-$0.000038 & 0.000630 \\
2454730.556153 &    0.000084 & 0.000629 \\
2454730.557982 & $-$0.000347 & 0.000628 \\
2454730.559811 & $-$0.000330 & 0.000627 \\
2454730.561651 & $-$0.000274 & 0.000625 \\
\hline \hline \end{tabular} \end{table}


\section{Light curve modelling}                                                           \label{sec:lc}

\citet{Me08mn,Me09mn} presented an homogeneous analysis of fourteen well-observed TEPs. We have used the same methods here, so our findings are fully compatible with the results in these papers. The analysis splits naturally into two steps, the first of which is modelling the light curves \citep{Me08mn}, and the second of which is including additional observed quantities to determine the physical properties of the star and planet \citep{Me09mn}.

The transit light curves of WASP-5 were modelled using the {\sc jktebop}%
\footnote{{\sc jktebop} is written in {\sc fortran77} and the source code is available at {\tt http://www.astro.keele.ac.uk/$\sim$jkt/}}
code \citep{Me++04mn,Me++04mn2}, which is based on the {\sc ebop} program originally developed for eclipsing binary star systems \citep{PopperEtzel81aj,Etzel81conf,NelsonDavis72apj}. The parameters of the fit include the fractional radii of the star and planet, which are directly dependent on the shape of the light curve and are defined to be $r_{\rm A} = \frac{R_{\rm A}}{a}$ and $r_{\rm b} = \frac{R_{\rm b}}{a}$ where $R_{\rm A}$ and $R_{\rm b}$ are the absolute radii of the components and $a$ is the orbital semimajor axis. The fractional radii are actually parameterised as $r_{\rm A}+r_{\rm b}$ and $k = \frac{r_{\rm A}}{r_{\rm b}}$ in {\sc jktebop}, as these are only weakly correlated with each other \citep{Me08mn}.  The other `shape' parameters are the orbital inclination, $i$, and the limb darkening coefficients of the star, $u_{\rm A}$ and $v_{\rm A}$ (see below). The mass ratio of the system is also a parameter as it affects the shapes of the components. We adopted a value of 0.00155, and verified that large changes in this number had a negligible effect on our results.

\reff{The main analysis of our photometry was undertaken before any transit timings were available in the literature. We therefore allowed the midpoints of the two eclipses to float freely, by including the orbital period and eclipse midpoint as fitted parameters, to guard against possible period. We have also modelled the two light curves individually to measure two times of minimum light: $2454707.82388 \pm 0.00029$ and $2454730.62115 \pm 0.00032$ (HJD). During this process we found that the error bars in our data (which come from the {\sc idl}/{\sc aper} routine) are slightly conservative and result in reduced $\chi^2$ values of $\chi^{\ 2}_{\rm red} = 0.95$ and $0.75$ for the two datasets. We therefore multiplied the error bars by $\sqrt{\chi^{\ 2}_{\rm red}}$ before combining the data from the two nights into one light curve \citep[see e.g.][]{Bruntt+06aa}.}

\reff{Our two times of minimum light have been augmented with the four transit times given by \citet{Gillon+08xxx} and used to determine a linear ephemeris. \citeauthor{Gillon+08xxx} found that a straight line was not a good representation of the data ($\chi^{\ 2}_{\rm red} = 5.7$). With our additional data we can associate this discrepancy with the transit time from their FTS light curve, which is earlier by 3$\sigma$ that predicted by our linear ephemeris. Further data are needed to decide whether the cause is astrophysical or observational. Using all times of minimum light, we find
$$ T_0 = {\rm HJD} 2\,454\,375.62494 (24) + 1.6284246 (13) \times E $$
where quantities in parentheses denote the uncertainty in the final digit of the preceding number.}

\begin{figure} \includegraphics[width=0.48\textwidth,angle=0]{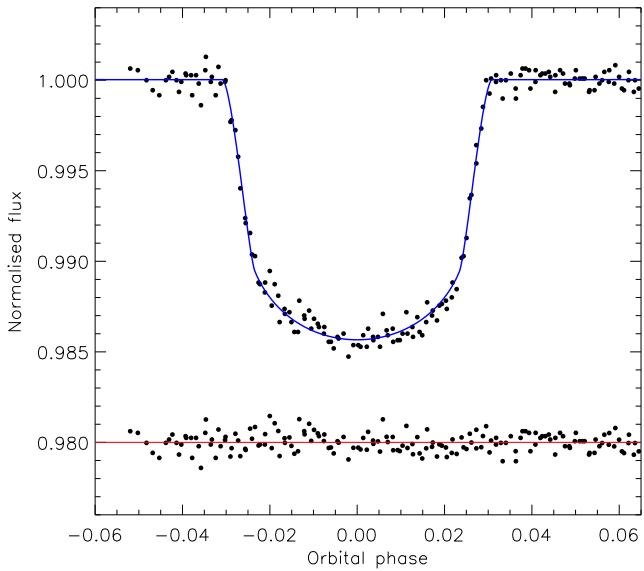}
\caption{\label{fig:lcfit} Phased light curve of the two transits of WASP-5
compared to the best fit found using {\sc jktebop} and the quadratic LD law
with both LD coefficients included as fitted parameters. The residuals of the
fit are plotted at the bottom of the figure, offset from zero.} \end{figure}

\begin{table*} \caption{\label{tab:lcfit1} Parameters of the {\sc jktebop}
best fits of the light curve of WASP-5 for different LD laws with the
coefficients fixed at theoretically predicted values. For each part of the
table the upper quantities are fitted parameters and the lower quantities
are derived parameters. Results were not calculated using the cubic LD law
because theoretical LD coefficients are not available.}
\begin{tabular}{l r@{\,$\pm$\,}l r@{\,$\pm$\,}l r@{\,$\pm$\,}l r@{\,$\pm$\,}l r@{\,$\pm$\,}l}
\hline \hline
\                     &\mc{Linear LD law}&\mc{Quadratic LD law}&\mc{Square-root LD law}&\mc{Logarithmic LD law} \\
\hline
$r_{\rm A}+r_{\rm b}$ &   0.2036  & 0.0056  &   0.2009  & 0.0057  &   0.2085  & 0.0050  &   0.2024  & 0.0057   \\
$k$                   &   0.1103  & 0.0010  &   0.1096  & 0.0010  &   0.1119  & 0.0007  &   0.1100  & 0.0009   \\
$i$ (deg.)            &  86.17    & 0.84    &  86.66    & 0.96    &  85.37    & 0.60    &  86.37    & 0.90     \\
$u_{\rm A}$           &   \mc{0.60 fixed}   &  \mc{0.40 fixed}    &   \mc{0.20 fixed}   &    \mc{0.70 fixed}   \\
$v_{\rm A}$           &        \mc{ }       &  \mc{0.30 fixed}    &   \mc{0.50 fixed}   &    \mc{0.20 fixed}   \\
\hline
$r_{\rm A}$           &   0.1834  & 0.0048  &   0.1811  & 0.0051  &   0.1875  & 0.0044  &   0.1824  & 0.0050   \\
$r_{\rm b}$           &   0.02022 & 0.00071 &   0.01984 & 0.00070 &   0.02098 & 0.00061 &   0.02006 & 0.00071  \\
$\sigma$ ($m$mag)     &     \mc{0.5696}     &     \mc{0.5593}     &     \mc{0.5586}     &     \mc{0.5602}      \\
\hline\hline \end{tabular} \end{table*}

\begin{table*} \caption{\label{tab:lcfit2} Parameters of the {\sc jktebop}
best fits of the light curve of WASP-5 for different LD laws with the linear
coefficients included as fitted parameters and the nonlinear coefficients
fixed at theoretically predicted values which are perturbed in the Monte
Carlo error analysis. For each part of the table the upper quantities are
fitted parameters and the lower quantities are derived parameters.}
\begin{tabular}{l r@{\,$\pm$\,}l r@{\,$\pm$\,}l r@{\,$\pm$\,}l r@{\,$\pm$\,}l r@{\,$\pm$\,}l}
\hline \hline
\                     & \mc{Linear LD law} &\mc{Quadratic LD law}&\mc{Square-root LD law}&\mc{Logarithmic LD law}& \mc{Cubic LD law} \\
\hline
$r_{\rm A}+r_{\rm b}$ &  0.2097  & 0.0052  &   0.2034  & 0.0062  &    0.2058  & 0.0053   &    0.2054  & 0.0056   &  0.2059  & 0.0060 \\
$k$                   &  0.1125  & 0.0009  &   0.1102  & 0.0013  &    0.1112  & 0.0011   &    0.1110  & 0.0011   &  0.1114  & 0.0012  \\
$i$ (deg.)            & 85.16    & 0.64    &  86.22    & 1.02    &   85.77    & 0.75     &   85.83    & 0.77     & 85.73    & 0.81   \\
$u_{\rm A}$           &  0.508   & 0.031   &   0.378   & 0.048   &    0.227   & 0.049    &    0.659   & 0.058    &  0.495   & 0.037  \\
$v_{\rm A}$           &      \mc{ }        & \mc{0.30 perturbed} &  \mc{0.50 perturbed}  &  \mc{0.20 perturbed}  &\mc{0.10 perturbed}\\
\hline
$r_{\rm A}$           &  0.1885  & 0.0046  &   0.1832  & 0.0055  &    0.1852  & 0.0047   &    0.1849  & 0.0049   &  0.1852  & 0.0052 \\
$r_{\rm b}$           &  0.02119 & 0.00067 &   0.02019 & 0.00083 &    0.02058 & 0.00070  &    0.02053 & 0.00071  &  0.02063 & 0.00077\\
$\sigma$ ($m$mag)     &    \mc{0.5576}     &     \mc{0.5579}     &      \mc{0.5578}      &      \mc{0.5579}      &    \mc{0.5578}    \\
\hline \hline \end{tabular} \end{table*}

In \citet{Me08mn} it was found that the treatment of the stellar limb darkening (LD) requires careful thought. LD is included in {\sc jktebop} as a choice of several parametric `laws' for which coefficients must be specified. The linear law is inadequate for high-precision observations, whereas more sophisticated nonlinear laws suffer from strong correlations between their coefficients \citep{Me++07aa,Me08mn}. Furthermore, theoretically calculated LD coefficients depend on stellar model atmospheres and do not match the highest-quality observations. The solution is to adopt nonlinear laws but to fix one of the coefficients at a reasonable value. An alternative would be to re-parameterise the two-coefficient laws so the coefficients have weaker correlations, but this has little effect on the other parameters of the fit.

We have incorporated LD using the linear, quadratic, square-root, logarithmic or cubic laws. We present one set of solutions where both the linear ($u_{\rm A}$) and nonlinear ($v_{\rm A}$) LD coefficients are fixed to theoretical values (Table\,\ref{tab:lcfit1}). We also give a set of solutions where $u_{\rm A}$ is a fitted parameter and $v_{\rm A}$ is fixed to theoretical values (Table\,\ref{tab:lcfit2}). In the second case, $v_{\rm A}$ is perturbed by $\pm$0.1 on a flat distribution during the error analysis. The theoretical LD coefficients are taken to be the approximate mean of values found by bilinear interpolation within the tabulations of \citet{Vanhamme93aj}, \citet{Claret00aa,Claret04aa2} and \citet{ClaretHauschildt03aa}.

We have used Monte Carlo simulations to assign uncertainties to the parameters found from the light curve, following the procedures of \citet{Me++04mn2,Me+05mn}. Each fit was followed by 1000 Monte Carlo simulations and 1$\sigma$ error bars were calculated from the scatter in the simulation results (Tables \ref{tab:lcfit1} and \ref{tab:lcfit2}). This does not account for any systematic errors which might be in our light curves, so this possibility was investigated using the residual-permutation approach \citep{Jenkins++02apj} and found to be marginally significant. The residuals in Fig.\,\ref{fig:lcfit} suggest there are minor systematic effects, in agreement with the residual-permutation results.

From Tables \ref{tab:lcfit1} and \ref{tab:lcfit2} it can be seen that the solutions with LD coefficients fixed at theoretical values have slightly larger residuals than those where the linear LD coefficient is a parameter of the fit. We have therefore used the latter solutions for our final results, combined following the methods of \citet{Me08mn} and incorporating the systematic uncertainties found with the residual-bead method. The final values of the photometric parameters are given in Table\,\ref{tab:lcfinal}. They are in agreement with and more precise than the values found in the two previous studies of WASP-5, \citet{Anderson+08mn} and \citet{Gillon+08xxx}, which used the same photometric data. Including the stellar velocity amplitude given by \citet{Anderson+08mn}, we measure the surface gravity of the planet to be $g_{\rm b} = 29.6 \pm 2.8$\mss\ \citep[see][]{Me++07mn}.

\begin{table*} \centering \caption{\label{tab:lcfinal} Final parameters of the fit to the
light curve of WASP-5, based on the individual solutions given in Table\,\ref{tab:lcfit2}.
The orbital ephemeris was determined separately (Section\,\ref{sec:lc}). The results of
\citet{Anderson+08mn} and \citet{Gillon+08xxx}, which are based on the same photometric
data, are included for comparison, and are without error bars if they were not directly
quoted quantities. $T_0$ is the reference time of mid-transit.}
\setlength{\tabcolsep}{4.5pt}
\begin{tabular}{l r@{\,$\pm$\,}l r@{\,$\pm$\,}l r@{\,$\pm$\,}l}
\hline \hline
\                     &    \mc{This work}       & \mc{\citet{Anderson+08mn}}            & \mc{\citet{Gillon+08xxx}}              \\
\hline
Orbital period (d)    &    1.6284246 & 0.0000013& \erm{1.6284296}{0.0000048}{0.0000037} & \erm{1.6284279}{0.0000022}{0.0000049}  \\
$T_0$ (HJD)           &2454375.62494 & 0.00024  & \erm{2454375.62466}{0.00026}{0.00025} & \erm{2454373.99598}{0.00025}{0.00019}  \\
$r_{\rm A}+r_{\rm b}$ &      0.2052  & 0.0068   & \mc{0.197}                            & \mc{0.198}                             \\
$k$                   &      0.1110  & 0.0014   & \erm{0.1092}{0.0030}{0.0017}          & \erm{0.1086}{0.0010}{0.0013}           \\
$i$ ($^\circ$)        &     85.8     & 1.1      & \mc{85.0 to 90.0}                     & \erm{86.9}{2.8}{0.7}                   \\
$r_{\rm A}$           &      0.1847  & 0.0061   & \mc{0.178}                            & \mc{0.179}                             \\
$r_{\rm b}$           &      0.02050 & 0.00091  & \mc{0.0194}                           & \mc{0.0194}                            \\
\hline \hline \end{tabular} \end{table*}


\section{Physical properties of WASP-5}                                                     \label{sec:absdim}

The photometric parameters determined above have been used to infer the physical properties of the WASP-5 system using the method outlined in \citet{Me09mn}. It is not possible to calculate the properties directly from observations, as one piece of information is missing and must be filled in by additional constraints. This approach requires as input the measured values of $r_{\rm A}$, $r_{\rm b}$, $i$, orbital period, and the stellar effective temperature, metallicity and velocity amplitude. For the latter three quantities we adopted $\Teff = 5700 \pm 100$\,K and $\MoH = 0.09 \pm 0.09$\,dex from \citet{Gillon+08xxx} and $K_{\rm A} = 278 \pm 8$\kms\ from \citet{Anderson+08mn}. We then interpolated within the tabulated predictions of several sets of theoretical stellar models to provide the optimum fit to the input quantities. The stellar models used were {\sf Claret} \citep{Claret04aa,Claret05aa,Claret06aa2,Claret07aa2}, {\sf Y$^2$} \citep{Demarque+04apjs} and {\sf Padova} \citep{Girardi+00aas}. The uncertainties in the input parameters were propagated into the output physical properties using a perturbation analysis \citep{Me09mn,Me++05aa}, which allows a detailed error budget to be obtained for each output quantity.

The use of theoretical stellar model predictions means that the calculated physical properties of the WASP-5 system are subject to systematic errors caused by any deviations of the models from reality. Unfortunately, stellar models are persistently unable to match the directly measured physical properties of low-mass eclipsing binary star systems, although this effect is likely due to magnetic activity \citep{Hoxie73aa,Clausen98conf,TorresRibas02apj,Ribas+08conf}. A lower limit can be set on the systematic errors by considering the variation in results from different sets of stellar models. This is only a lower limit because independent models have some physical ingredients in common, for example opacity tables. A probable upper limit can be set by avoiding stellar models and instead constraining the properties using an empirical main-sequence mass--radius relation for the star \citep{Me09mn}. Table\,\ref{tab:absdimall} contains the physical properties of WASP-5 calculated using the three different sets of stellar models and with the mass--radius relation.

\begin{table*}
\caption{\label{tab:absdimall} Physical properties for WASP-5, derived using either an
empirical stellar mass--radius relation or the predictions of different sets of stellar
evolutionary models. The stellar mass, radius, surface gravity and density are denoted
by $M_{\rm A}$, $R_{\rm A}$, $\log g_{\rm A}$ and $\rho_{\rm A}$, respectively.}
\begin{tabular}{l l r@{\,$\pm$\,}l r@{\,$\pm$\,}l r@{\,$\pm$\,}l r@{\,$\pm$\,}l r@{\,$\pm$\,}l r@{\,$\pm$\,}l}
\hline \hline
\ & \ & \mc{Mass--radius} & \mc{{\sf Padova} models}  & \mc{{\sf Y$^2$} models} & \mc{{\sf Claret} models} \\
\hline
$a$ & (AU)               & 0.02777& 0.00078& 0.02702&0.00046 & 0.02728&0.00043 & 0.02729&0.00049 \\[2pt]
$M_{\rm A}$ & (\Msun)    &  1.076 & 0.091  &  0.991 & 0.051  &  1.020 & 0.048  &  1.021 & 0.055  \\
$R_{\rm A}$ & (\Rsun)    &  1.103 & 0.060  &  1.073 & 0.039  &  1.083 & 0.040  &  1.084 & 0.039  \\
$\log g_{\rm A}$ & [cgs] &  4.385 & 0.024  &  4.373 & 0.030  &  4.377 & 0.029  &  4.377 & 0.030  \\[2pt]
$\rho_{\rm A}$ & (\psun) &  0.803 & 0.080  &  0.803 & 0.080  &  0.803 & 0.080  &  0.803 & 0.080  \\[2pt]
$M_{\rm b}$ & (\Mjup)    &  1.70  & 0.11   &  1.604 & 0.073  &  1.635 & 0.070  &  1.637 & 0.075  \\
$R_{\rm b}$ & (\Rjup)    &  1.191 & 0.063  &  1.159 & 0.055  &  1.170 & 0.055  &  1.171 & 0.056  \\
$g_{\rm b}$ & (\mss)     &   29.6 & 2.7    &   29.6 & 2.8    &   29.6 & 2.8    &   29.6 & 2.8    \\
$\rho_{\rm b}$ & (\pjup) &  1.00  & 0.14   &  1.03  & 0.14   &  1.02  & 0.14   &   1.02 & 0.14   \\[2pt]
Age & (Gyr)              & \mc{ } & \erm{6.6}{1.9}{3.3} & \erm{5.9}{2.2}{1.9} & \erm{6.8}{3.0}{2.5} \\
\hline \hline \end{tabular} \end{table*}

The properties calculated using stellar models are in good agreement with each other, whereas those determined using the mass--radius relation are offset to larger values. The difference is most significant for the orbital semimajor axis ($a$) and the stellar mass ($M_{\rm A}$) as these quantities are quite affected by systematics but almost unaffected by observational uncertainties. The error budget for the output parameters is given in Table\,\ref{tab:err}, and shows that more precise stellar \Teff\ and \MoH\ measurements would be useful, as well as further high-precision photometry.

\begin{table} \centering
\caption{\label{tab:err} Detailed error budget for the calculation of the physical
properties of the WASP-5 system from the light curve parameters, stellar velocity
amplitude, and the predictions of the {\sf Claret} stellar models. Each number in
the table is the fractional contribution to the final uncertainty of an output
parameter from the error bar of an input parameter. The final uncertainty for each
output parameter (not given) is the quadrature sum of the individual contributions
from each input parameter. $P$ represents the orbital period.}
\begin{tabular}{l|rrrrrrrr}
\hline \hline
Output            & \multicolumn{7}{c}{Input parameter} \\
parameter         & $P$   & $K_{\rm A}$ & $i$   & $r_{\rm A}$ & $r_{\rm b}$ & \Teff & \MoH  \\
\hline
Age               &       &             &       &    0.341    &             & 0.753 & 0.556 \\
$a$               &       &    0.001    &       &    0.075    &             & 0.760 & 0.644 \\[2pt]
$M_{\rm A}$       &       &             &       &    0.075    &             & 0.760 & 0.645 \\
$R_{\rm A}$       &       &             &       &    0.871    &             & 0.374 & 0.317 \\
$\log g_{\rm A}$  &       &             &       &    0.967    &             & 0.195 & 0.166 \\
$\rho_{\rm A}$    &       &             &       &    1.000    &             &       &       \\[2pt]
$M_{\rm b}$       &       &    0.627    & 0.031 &    0.059    &             & 0.592 & 0.502 \\
$R_{\rm b}$       &       &             &       &    0.028    &    0.927    & 0.284 & 0.241 \\
$g_{\rm b}$       &       &    0.307    & 0.015 &             &    0.951    &       &       \\
$\rho_{\rm b}$    &       &    0.208    & 0.010 &    0.010    &    0.969    & 0.099 & 0.083 \\
\hline \hline
\end{tabular} \end{table}

Table\,\ref{tab:absdim} gives the final physical properties for WASP-5, with statistical errors obtained from the perturbation analysis and systematic errors from consideration of the interagreement between results calculated using the three different sets of stellar evolutionary models. We have adopted the values of the properties obtained using the {\sf Claret} models, in order to retain homogeneity with the analyses of \citet{Me09mn}. A comparison with the results of \citet{Anderson+08mn} and \citet{Gillon+08xxx} shows that all quantities agree within the quoted uncertainties. The main difference is that our more extensive photometry yields a larger radius and thus a lower density for the planet; this arises from the light curve analysis via the parameter $r_{\rm b}$.

\begin{table*}
\caption{\label{tab:absdim} Final physical properties for WASP-5. The first error
bars are statistical and the second are systematic. The corresponding results from
\citet{Anderson+08mn} and \citet{Gillon+08xxx} have been included for comparison.}
\begin{tabular}{l l r@{\,$\pm$\,}c@{\,$\pm$\,}l c c}
\hline \hline
    &     & \multicolumn{3}{c}{Final result (this work)} & \citet{Anderson+08mn}          & \citet{Gillon+08xxx} \\
\hline
$a$ & (AU)               &  0.02729 & 0.00049 & 0.00027  & \er{0.02683}{0.00088}{0.00075} & \er{0.0267}{0.0012}{0.0008} \\
$M_{\rm A}$ & (\Msun)    &  1.021   & 0.055   & 0.030    & \er{0.972}{0.099}{0.079}       & \er{0.96}{0.13}{0.09} \\
$R_{\rm A}$ & (\Rsun)    &  1.084   & 0.040   & 0.011    & \er{1.026}{0.073}{0.044}       & \er{1.029}{0.056}{0.069} \\
$\log g_{\rm A}$ & [cgs] &  4.377   & 0.030   & 0.004    & \er{4.403}{0.039}{0.048}       & \er{4.395}{0.043}{0.040} \\
$\rho_{\rm A}$ & (\psun) &  0.803   & 0.080   & 0.000    & 0.90                           & $0.88 \pm 0.12$ \\
$M_{\rm b}$ & (\Mjup)    &  1.637   & 0.075   & 0.033    & \er{1.58}{0.13}{0.08}          & \er{1.58}{0.13}{0.10} \\
$R_{\rm b}$ & (\Rjup)    &  1.171   & 0.056   & 0.012    & \er{1.090}{0.094}{0.058}       & \er{1.087}{0.068}{0.071} \\
$g_{\rm b}$ & (\mss)     & 29.6 & \multicolumn{2}{l}{2.8}& \er{30.5}{3.2}{4.1}            & \er{30.5}{4.0}{2.9} \\
$\rho_{\rm b}$ & (\pjup) &  1.02    & 0.14    & 0.01     & \er{1.22}{0.19}{0.24}          & \er{1.23}{0.26}{0.16} \\
\hline \hline
\end{tabular} \end{table*}


\section{What is the optimal exposure time?}                                                    \label{sec:sn}

Having presented and interpreted our observations, we now present detailed signal to noise (S/N) calculations which can be used to plan similar projects. The output quantities include the amount of defocussing required, and the total noise to signal ratio in millimagnitudes. The tractability of these calculations demands several approximations, so the results should not be overinterpreted. The {\sc idl} code used for the calculations can be obtained from the first author.

The important point here is that the quantity of interest for a given observing sequence is the achievable S/N {\em per unit time} rather than per observation. This is limited in particular by CCD readout for short exposures and sky background noise for long integrations. There will be an optimum value between these two extremes, which depends on the target and observing facilities.

We start with count rates for the target (overall) and for the sky background (per pixel) gathered from the {\sc signal}%
\footnote{Information on the Isaac Newton Group's {\sc signal} code can be found at {\tt http://catserver.ing.iac.es/signal/}}
code. These are appropriate for the Isaac Newton Telescope and Wide Field Camera, so must be scaled for the difference in telescope aperture (1.5\,m versus 2.5\,m) and plate scale (0.39\as\,pixel$^{-1}$ instead of 0.33\as\,pixel$^{-1}$). We adopt the default values for sky brightness in dark, grey and bright time; this approximation can be improved as and when additional information is available. The count rates are specified in electrons, so the CCD gain level is taken into account with the input parameters rather than as part of the S/N calculations.

For a given $t_{\rm exp}$ (in seconds) we calculate the total electrons obtained from the target, and the associated Poisson noise, to be
\begin{equation} S_{\rm target} = t_{\rm exp} C_{\rm target} \end{equation}
\begin{equation} N_{\rm target} = \sqrt{t_{\rm exp} C_{\rm target}} \end{equation}
where $t_{\rm exp}$ is the exposure time and $C_{\rm target}$ is the electrons per second from the target. In order to estimate the amount of defocussing required (the actual shape of the PSF does not enter these calculations), we distribute these equally over $n_{\rm pix}$ pixels by specifying a maximum number of electrons per pixel from the target and the background, $m_{\rm total}$. Fig.\,\ref{fig:psf} shows that this approach is imperfect so we also specify the maximum electrons per pixel from the target alone, $m_{\rm target}$. This allows us to limit the {\em range} of counts within a structured PSF, and thus avoid saturating the most illuminated pixels.

If the sky background is low then we need only to account for the limit set by $m_{\rm target}$ in working out how many pixels to put the target electrons into:
\begin{equation} n_{\rm pix} (m_{\rm target}) = \frac{S_{\rm target}}{m_{\rm target}} \end{equation}
If the sky background is high then we must increase the number of pixels containing counts from the target to avoid broaching the $m_{\rm total}$ limit. In this case the number of pixels in the PSF is:
\begin{equation} n_{\rm pix} (m_{\rm total}) = \frac{S_{\rm target}}{m_{\rm total}-C_{\rm sky}t_{\rm exp}} \end{equation}
where $C_{\rm sky}$ is the electrons per second per pixel from the sky background. For a specific situation, $n_{\rm pix}$ is taken to be the larger of the two alternatives.

Now we have found $n_{\rm pix}$ we can calculate the total sky level in the PSF, and the ensuing Poisson noise, in one observation:
\begin{equation} S_{\rm sky} = t_{\rm exp} n_{\rm pix} C_{\rm sky} \end{equation}
\begin{equation} N_{\rm sky} = \sqrt{t_{\rm exp} n_{\rm pix} C_{\rm sky}} \end{equation}
The contribution from CCD detector readout noise is:
\begin{equation} N_{\rm ron} = n_{\rm ron} \sqrt{n_{\rm pix}} \end{equation}
where $n_{\rm ron}$ is the readout noise per pixel in electrons.

The treatment of flat-field noise is more difficult, as the PSFs of one star in different observations will have many pixels in common whose flat-field noise will therefore cancel out. We approximate this noise contribution using $f_{\rm flat}$, the flat-field noise per pixel (expressed as a fraction of the electrons in a pixel), and apply it to a number $n_{\rm flat}$ pixels where $n_{\rm flat} \leqslant n_{\rm pix}$. The flat-field noise is then
\begin{equation} N_{\rm flat} = f_{\rm flat} \left(\frac{S_{\rm target}}{n_{\rm pix}}+C_{\rm sky}\right) \frac{n_{\rm flat}}{n_{\rm pix}} \end{equation}

We can now express these noise levels as fractions of the total electrons from the star in magnitude units:
\begin{equation} \sigma_{\rm target} = -2.5 \log_{10} \left(\frac{S_{\rm target}-N_{\rm target}}{S_{\rm target}}\right) \end{equation}
\begin{equation} \sigma_{\rm sky} = -2.5 \log_{10} \left(\frac{S_{\rm target}-N_{\rm sky}}{S_{\rm target}}\right) \end{equation}
\begin{equation} \sigma_{\rm ron} = -2.5 \log_{10} \left(\frac{S_{\rm target}-N_{\rm ron}}{S_{\rm target}}\right) \end{equation}
\begin{equation} \sigma_{\rm flat} = -2.5 \log_{10} \left(\frac{S_{\rm target}-N_{\rm flat}}{S_{\rm target}}\right) \end{equation}
The noise due to scintillation can be found using the relation given by \citet{Dravins+98pasp} and originally due to \citet{Young67aj}:
\begin{equation} \sigma_{\rm scint} = 0.004 D^{-2/3} X^{7/4} e^{-h/H} (2t_{\rm exp})^{-1/2} \end{equation}
where $D$ is the telescope aperture (m), $X$ is the airmass, $h$ is the altitude of the telescope (m), and $H = 8000$\,m is the scale height of the atmosphere. Note that scintillation varies seasonally, and the exponent of the airmass term actually varies between 1.5 and 2.0 depending on the direction of the wind relative to the telescope pointing \citep{Dravins+98pasp}. These caveats illustrate the approximate nature of S/N calculations.

We now have five noise contributions which can be added in quadrature to reveal the total noise for one magnitude measurement in one observation:
\begin{equation} \sigma_{\rm total} = \sqrt{\sigma_{\rm target}^2 + \sigma_{\rm sky}^2 + \sigma_{\rm ron}^2 + \sigma_{\rm flat}^2 + \sigma_{\rm scint}^2} \end{equation}
The ratio of noise to signal per unit time is then
\begin{equation} \rho_{\rm total} = \sigma_{\rm total} \sqrt{t_{\rm exp}+d_{\rm readout}} \end{equation}
where $d_{\rm readout}$ is the total dead time per observation (normally dominated by the CCD readout time). In reality there will be additional noise in differential photometry due to measurement of the magnitude of the comparison star. We do not include this here as its contribution to the total noise can be driven to a much smaller value than for other sources by using many comparison stars.

Using the above formulae we have performed S/N calculations for the observations presented in this work. We adopted $m_{\rm target} = 10\,000$, $M_{\rm total} = 40\,000$, $n_{\rm ron} = 4.0$, $X = 1.2$, and neglected flat-fielding noise. We set $d_{\rm readout} = 91$\,s to represent the first of our two observing sequences. The predicted $\sigma_{\rm total}$ for $t_{\rm exp} = 120$\,s is 0.54\,mmag (dark time), as compared to the 0.50\,mmag we actually achieved. This good agreement could be further improved by tinkering with the input parameters to the S/N calculations.

\begin{figure} \includegraphics[width=0.48\textwidth,angle=0]{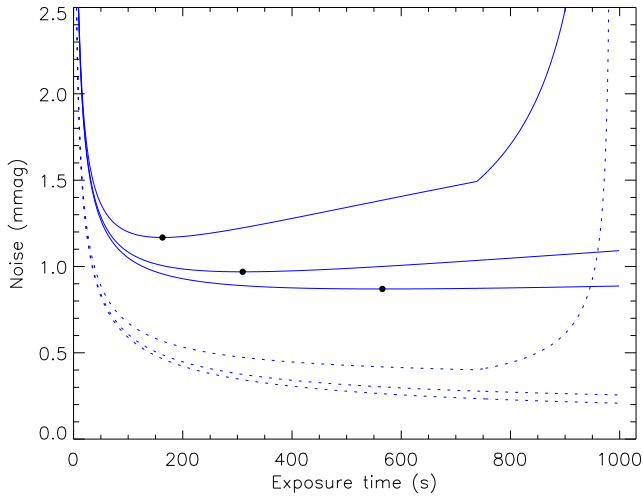}
\caption{\label{fig:sn} Predicted noise levels for the observations
presented in this work, but as a function of exposure time. The dotted
curves show the predicted noise level per observation for dark, grey and
bright time (lower to upper curves). The solid curves show the predicted
noise level per minute of telescope time (dark, grey, bright), and the
optimum exposure times are indicated with filled circles.} \end{figure}

The resulting curves of noise level per observation and per minute are plotted as a function of $t_{\rm exp}$ in Fig.\,\ref{fig:sn}. The `sweet spots' representing the lowest achievable noise level per minute (indicated by filled circles) occur at $t_{\rm exp} = 565$\,s (dark time), 310\,s (grey) and 163\,s (bright). Specifying a shorter readout time or fainter target star would shift the sweet spots towards smaller $t_{\rm exp}$. The abrupt change of slope visible in Fig.\,\ref{fig:sn} at $t_{\rm exp} = 740$\,s (bright time) occurs when the sky background becomes so high that the starlight has to be spread over extra pixels to avoid transgressing the $m_{\rm total}$ limit. Using these S/N calculations we find that, except for a target star which is very faint, defocussing is always the best approach.

The exposure times yielding the lowest noise per minute for the current situation are too long to sample the light variations through a transit of WASP-5 adequately, which is why used $t_{\rm exp} \leqslant 120$\,s. This is partly due to the long readout time of the DFOSC CCD: adopting $d_{\rm readout} = 39$\,s (appropriate for our second transit sequence) shortens the optimal exposure times to 371\,s (dark), 203\,s (grey) and 107\,s (bright).

\section{Discussion and conclusions} \label{sec:discussion}

We have presented high-precision light curves of two transits of the transiting extrasolar planetary system WASP-5, obtained by heavily defocussing the Danish 1.5\,m telescope at ESO La Silla. The defocussing caused the telescope PSF to take an approximately annular shape of diameter 40 pixels (16\as). Data reduction was undertaken using standard aperture photometry techniques. Debiassing and flat-fielding the raw CCD images improved the results but only by a small amount, probably due to the use of autoguiding to confine the stellar PSFs to the same pixels on each image. Our final light curves have a scatter around the best-fitting model of 0.50\,mmag and 0.59\,mmag for the two transits, observed in dark and grey time, respectively.

\reff{We have presented detailed signal to noise calculations for defocussed photometry, taking into account Poisson noise from the target and background, readout noise, scintillation, and flat-fielding noise. Application of the resulting formulae to the current case yields a predicted noise level of 0.54\,mmag (dark time), which is in good agreement with the actual values. We use these calculations to assess what exposure times would give the best overall signal to noise per unit time, finding 565\,s (dark time) to 163\,s (bright time). Such large values arise partly because of the long readout time of the DFOSC CCD, and are impractical because they would not properly sample the light variations through transit.}

\reff{We find that telescope defocussing is a viable way of obtaining high-quality light curves of point sources over short periods of time. Its main advantages are that flat-fielding noise is averaged down and a smaller proportion of the observing time spent reading out the CCD. The scintillation and Poisson noise levels per unit time are ultimately limited by the telescope aperture rather than the observational strategy. One possible disadvantage of our approach is the use of a focal-reducing imager. These can be prone to internal reflections, which may have slightly increased the noise levels we find.} One definite advantage, though, is that the telescope was equatorially mounted, so the target and comparison stars could be kept to the same light path whilst observing a transit. It is not possible to do this with altitude-azimuth telescopes, even if an image derotator is used, making such observations liable to problems with systematic errors.

The resulting light curves were modelled with the {\sc jktebop} code using the approach outlined by \citet{Me08mn}. Limb darkening of the star was dealt with using five different LD laws, with similar results when one or both of the coefficients were included as parameters of the fit. We then adopted the observed \Teff, \MoH\ and velocity amplitude of the star \citep{Anderson+08mn,Gillon+08xxx} and interpolated within the tabulated predictions of three different sets of theoretical stellar evolutionary models to derive the physical properties of the WASP-5 system \citep{Me09mn}. This approach has allowed us to assign both statistical and systematic errors to all output quantities. Our results are fully consistent with the homogeneous analyses of 14 transiting extrasolar planetary systems presented by \citet{Me08mn,Me09mn}.

We find that the planet WASP-5\,b has a density equal to that of Jupiter, and that the system has an age of approximately 6\,Gyr. The theoretical models presented by \citet{Fortney++07apj} can match WASP-5\,b's mass and radius without requiring a heavy-element core. WASP-5 is also in line with the trend of planets with short orbital periods to have relatively high surface gravities \citep{Me++07mn} and masses \citep{Mazeh++05mn}. We calculate that its equilibrium temperature is $T_{\rm eq} = 1732 \pm 80$\,K, which may be high enough for WASP-5\,b to fall into the `pM class' of planets with large atmospheric opacities due to oxides of titanium and vanadium \citep{Fortney+08apj}. Its \citet{Safronov72} number is $\Theta = 0.0755 \pm 0.0096$, which makes it a Class\,II planet according to the study of \citet{HansenBarman07apj}. However, \citet{Me09mn} found that the division between Class\,I and Class\,II objects is under threat from the steady addition of new discoveries to the catalogue of known transiting extrasolar planets, and may not be statistically significant. The high equilibrium temperature of WASP-5\,b makes the system a good candidate for detecting the secondary eclipse at phase 0.5 \citep[e.g.][]{Deming+05nat,Harrington+07nat,Knutson+07nat}, which would help our understanding of the physics of the atmospheres of gas giants. Such observations, as well as more precise measurements of the stellar velocity amplitude, \Teff\ and \MoH, are encouraged.


\section*{Acknowledgments}

The photometric observations presented in this work will be made available at the CDS ({\tt http://cdsweb.u-strasbg.fr/}) and at \\ {\tt http://www.astro.keele.ac.uk/$\sim$jkt/}.

We would like to thank the referee for a very useful report. JS acknowledges financial support from STFC in the form of a postdoctoral research assistant position. Astronomical research at the Armagh Observatory is funded by the Northern Ireland Department of Culture, Arts and Leisure (DCAL). VB, SCN, LM and GS acknowledge support by funds of Regione Campania, L.R. n.5/2002, year 2005 (run by Gaetano Scarpetta).

The following internet-based resources were used in research for this paper: the ESO Digitized Sky Survey; the NASA Astrophysics Data System; the SIMBAD database operated at CDS, Strasbourg, France; and the ar$\chi$iv scientific paper preprint service operated by Cornell University.


\bibliographystyle{mn_new}

\end{document}